\documentstyle[12pt,aasms4,epsfig,psfig]{article}

\def \sax {{\it Beppo}SAX } 
\def\gta{ \lower .75ex \hbox{$\sim$} \llap{\raise .27ex \hbox{$>$}} }
\def\lta{ \lower .75ex\hbox{$\sim$} \llap{\raise .27ex \hbox{$<$}} }

\begin{document}

\title{Spectral Energy Distributions of Flat Spectrum Radio Quasars observed
with $Beppo$SAX.}  

\author{Tavecchio F.$^1$, Maraschi L.$^1$, Ghisellini G.$^2$, Celotti
A.$^3$, Chiappetti L.$^4$, Comastri A.$^5$, Fossati G.$^6$, Grandi
P.$^7$, Pian E.$^8$, Tagliaferri G.$^2$, Treves A.$^9$, Sambruna R.$^{10}$}

\altaffiltext{1}{Osservatorio Astronomico di Brera, via Brera 28, 20121,
Milano, Italy.}

\altaffiltext{2}{Osservatorio Astronomico di Brera, via Bianchi 46, 23807 Merate, Italy}

\altaffiltext{3}{SISSA/ISAS, via Beirut 2-4, 34014 Trieste, Italy}

\altaffiltext{4}{IFC/CNR, via Bassini 15 , 20133, Milano,  Italy} 

\altaffiltext{5}{Osservatorio Astronomico di Bologna, via Ranzani 1, 40127, Bologna, Italy}

\altaffiltext{6}{CASS, University of California, San Diego, CA 92093-0424}

\altaffiltext{7}{IAS, IAS/CNR, via Fosso del Cavaliere, 00133 Roma, Italy}

\altaffiltext{8}{Osservatorio  Astronomico di Trieste, via Tiepolo 11,
24131 Trieste, Italy}

\altaffiltext{9}{Universita' dell'Insubria, via Valleggio 11, 22100, Como, Italy}

\altaffiltext{10}{George Mason University, 4400 University Dr., M/S 3F3,
Fairfax, VA, 22030-4444}

\setcounter{footnote}{0}

\begin{abstract}

We report the $Beppo$SAX observations of 6 Flat Spectrum Radio
Quasars. Three of them have a clear detection up to 100 keV with the PDS
instrument. For 4 objects the X-ray spectrum is satisfactorily fitted by
a power-law continuum with Galactic absorption. 2251+158 show the
presence of absorption higher than the galactic value, while the spectrum
of the source 0208-512 shows a complex structure, with evidence of
absorption at low energy. We construct the Spectral Energy Distributions
adding historical data to the broad band X-ray spectra obtained with
$Beppo$SAX and reproduce them with a one-zone Synchrotron-Inverse Compton
model (including both SSC and External Compton). The implications are
briefly discussed.
\end{abstract}


\section{Introduction}

Blazars are the best laboratory to study the physics of relativistic
jets. The non-thermal continuum amplified by relativistic beaming (e.g.,
Urry \& Padovani 1995), is a unique tool to probe the physical processes
acting in the jet.  In the last decade the interest for Blazars has been
renewed by the EGRET discovery of 66 Blazars (Hartman et al. 1999) as
strong $\gamma $-ray emitters. In several cases the $\gamma $-ray
emission dominates the apparent bolometric luminosity. Moreover the short
variability timescales, together with the absence of absorption of high
energy photons (through the $\gamma \gamma$-pair production), directly
imply that the $\gamma $-ray emission is also relativistically beamed
(e.g. Dondi \& Ghisellini 1995).

Despite the large variety of classifications, there is growing evidence
that Blazars form a single population. Their different spectral
properties can be unified within a spectral sequence in which the leading
parameter is the total luminosity (Fossati et al. 1998). At the
high-luminosity extreme of the sequence we find the Flat Spectrum Radio
Quasars, the most luminous blazars, characterized by the presence of
bright emission lines in the optical-UV spectrum and, in some cases,
strong {\it UV bumps}, signatures of the underlying accretion process
(e.g. Pian et al. 1999 and references therein). Most of the Blazars
detected by EGRET (Hartman et al. 1999) belong to this sub-class but not
all FSRQs with comparable SEDs have been detected in the $\gamma $-ray
domain.

In the study of FSRQs, observations in the X-ray band play a fundamental
role. The X-ray emission from FSRQ is dominated by the continuum
originating from Inverse Compton scattering between relativistic
electrons in the jet and the soft photons produced by the disk and/or in
the Broad Line Region (Dermer \& Schlickeiser 1993; Sikora, Begelman \&
Rees 1994). The possibility to measure the IC component yelds important
constraints on the emission models. Moreover the study of the soft X-ray
spectrum (below 1 keV) provides insight on the presence of intrinsic
absorption, on the possible contribution of the high energy end of the
synchrotron emission or on the minimum energy of emitting particles. The
latter point is particularly interesting because the determination of the
minimum energy of particles is a fundamental step in the estimate of the
total power transported by the jet (e.g. Celotti, Padovani \& Ghisellini
1997).

The {\it Beppo}SAX satellite has the unique capability of covering the
wide energy range $0.1- 200$ keV. For this reason it is ideal to study
FSRQs. In particular the knowledge of the spectrum in this band, together
with the $\gamma $-ray spectrum measured by EGRET, provide the most
complete spectral information on the high energy component (see e.g.,
Tavecchio et al. 2000, hereafter Paper I).  For these reasons we started
an observational program with {\it Beppo}SAX of a sample of FSRQs,
(containing a total of 50 sources) extracted from the 2-Jy sample by
Padovani \& Urry (1992). Using the flux threshold $F_{\rm 1\, keV}>0.5
\mu$Jy the number of sources reduces to 19. In Paper I we discussed the
case of three FSRQ detected by EGRET, namely 0836+710, 1510-089 and
2230+114. Here we report the analysis of $Beppo$SAX data of 6 other 
sources, 3 detected by EGRET and three without an EGRET detection. Adding
these 6 sources to the 3 FSRQs analyzed in Paper I and other 3 sources
(3C279, 0528+134, PKS 0537-441) discussed elsewhere (Maraschi et
al. 1998; Ghisellini et al. 1999, Pian et al., in prep) a total of 12
FSRQs (out of 19 with X-ray flux $F_{\rm 1\, keV}>0.5 \mu$Jy) from the 2
Jy sample were observed by $Beppo$SAX. The sources discussed here are
listed in Table 1. Partial and preliminary results were reported in
Maraschi \& Tavecchio (2001)

The paper is organized as follows: in section 2 we report the analysis of
{\it Beppo}SAX observations, in section 3 we discuss the emission models
for the SEDs and finally in Sect. 4 we discuss our results.

\section{{\it Beppo}SAX observations and Analysis}

The scientific payload of the $Beppo$SAX satellite (see Boella et
al. 1997) contains four coaligned Narrow Field Instruments (NFIs) and
two Wide Field Cameras. Two of the NFIs use concentrators to focalize
X-rays: the Low Energy Concentrator Spectrometer (LECS) has a detector
sensitive to soft-medium X-ray photons (0.1-10 keV), while the Medium
Concentrator Spectrometer (MECS) can detect photons in the medium
energy range 1.3-10 keV. The Phoswich Detector System (PDS), sensitive
from 12 up to 200 keV, consisting of four identical units, uses rocking
collimators so as to monitor source and background simultaneously with
interchangeable units. We will not be concerned here with the fourth NFI,
a High Pressure Gas Scintillation Proportional Counter (HPGSPC).

The {\it Beppo}SAX journal of observations is reported in Table 2, with
exposure times and observed count rates. None of the sources showed
significant flux variations during the observations. We therefore
obtained a cumulative spectrum for each source.

We analyzed the {\it Beppo}SAX spectral data using the standard software
packages XSELECT (v 1.4b) and XSPEC (v 11.0) and the September 97 version of
the calibration files released by the $Beppo$SAX Scientific Data Center
(SDC).  From the event files we extracted the LECS and MECS spectra in
circular regions centered around the source with radii of 8$^{\prime }$ and 
4$^{\prime }$ respectively (see the SAX Analysis Cookbook\footnote{\small
ftp://www.sdc.asi.it/pub/sax/doc/software\_docs/saxabc\_v1.2.ps.gz}).
The PDS spectra extracted with the standard pipeline with the rise-time
correction were directly provided by the $Beppo$SAX SDC. We used PDS
data rebinned with S/N$>3$. 

For the spectral analysis we considered the LECS data in the restricted
energy range 0.1-4 keV, because of known unsolved problems with the
response matrix at higher energies. Background spectra extracted from
blank field observations at the same position as the source were used.
We fitted rebinned LECS, MECS and PDS spectra jointly, allowing for two
variable different normalization factors to take into account
uncertainties in the intercalibration of different instruments (see SAX
Cookbook).

The first model chosen for the fitting procedure was in all cases a
single power law plus Galactic absorption. In the following we discuss in
detail the results for each source. The results of the spectral fits are
summarized in Table 3 and 4.

\subsection{0208-512}

The residuals of a simple power-law model to the LECS+MECS data show the
presence of a prominent absorption feature, located at $\sim 0.6$ keV,
(see Fig.1) and an excess at about 5 keV.  Inspection of the individual
MECS2 and MECS3 spectra reveals that the emission feature at 5 keV is
present only in the data obtained with the MECS3 indicating an
instrumental origin. Only the MECS2 data are used below.

The low energy feature can be modeled either with an absorption edge
($E=0.60^{0.75}_{0.33}$) or with an absorption trough
($E=0.89^{0.94}_{0.80}$) with comparable probability ($\chi^2 _r=0.72$
($P=0.72$) and $\chi ^2 _r=0.78$ ($P=0.75$), respectively ). However
these features find no obvious interpretation at the redshift of the
source. We then tried a warm absorber model (ABSORI of XSPEC), at the
redshift of the quasar, obtaining a satisfactory fit ($\chi
^2_r=0.89$). The fit requires a column density of the absorber equal to
$N_H=1.2^{10.3}_{1.2} \times 10^{22}$ cm$^{-2}$ and an ionization
parameter $\xi=262^{853}_{32}$, providing a reasonable physical
interpretation. The results of the fits are reported in Table 4.

Previous ROSAT (Sambruna 1997) and ASCA (Kubo et al. 1998) spectra show a
power-law X-ray spectrum with photon index $\Gamma _{2-10}=1.7\pm 0.1$,
similar to that found with \sax. Notably these observations (with
comparable flux) showed a smooth continuum, without any evidence for
spectral features.

The PDS/MECS normalization factor required by the data ($2.7\pm 1.3$) is
well above the 90\% upper limit (0.93) given by the SAX
Cookbook. 0208-512 is located in a rich field: although the LECS and MECS
images do not show other sources, a search using the NASA Extragalactic
Database\footnote{http://nedwww.ipac.caltech.edu/index.html} reveals the
presence of 13 sources catalogued as QSOs or X-ray sources, within a
circular region with radius 1 deg centered on our target. It is then
possible that the high-energy spectrum is contaminated by one or more
sources present in the FOV ($\sim 1$ deg) of the PDS.

In the fit presented here the PDS data are not included.

\subsection{0521-365}

The LECS+MECS spectrum is well fitted by an absorbed power-law
($\Gamma=1.77$ [1.73-1.85]) with column density consistent with the
Galactic value. Previous X-ray observations with EXOSAT ($\Gamma \sim
1.6$, Pian et al. 1996) and ROSAT ($\Gamma =1.92\pm 0.05$, Sambruna 1997)
are consistent with our results.

\subsection{1641+399 (3C345)}

A power law model ($\Gamma \simeq 1.6$) with Galactic absorption fits the
MECS+PDS data for $E>2$ keV up to $\sim200$ keV.

The inspection of the LECS image reveals the presence of several sources
(not present in the MECS image): in particular a low-luminosity Seyfert
galaxy, NGC 6212, located 5$^{\prime }$ away from the quasar hence within
the standard extraction region of the LECS data.  We conclude that the
LECS spectrum could be contaminated by the presence of other sources
close to 1641+399. In fact the residuals of a fit with an absorbed
power-law model of the LECS data alone ($\Gamma \simeq 1.8$, $\chi ^2
_r=2.7$) show large scatter around the model (Fig.3), with the presence
of possible features (with no straightforward interpretation) at energies
$\sim 2$ keV and below $\sim$1.0 keV. The features cannot simply be
modeled as due to absorption of the continuum by a warm absorber at the
redshift of the source.

A clear soft excess, interpreted as the hard tail of the UV bump, was
present in the ROSAT data discussed by Sambruna (1997).

\subsection{2223-052 (3C446)} 

An absorbed power-law with Galactic absorption provides a good fit to the
data. Our spectrum ($\Gamma\simeq 1.85$) is steeper than that measured on
one occasion by GINGA ($\Gamma =1.3$, Lawson et al. 1997). 
Moreover we do not confirm the ROSAT observations that seem to indicate an
absorption column higher than the Galactic value at the 99\% confidence
level (Sambruna 1997).

\subsection{2243-123}

A power-law with intermediate spectral index ($\Gamma=1.7$) fits well the
data.

The ROSAT observation reported by Siebert et al. (1998) showed a very
steep spectrum, with $\Gamma =2.9\pm 0.5$ and a flux about 2.5 times
larger than during our \sax pointing.  Given the large flux variation, the
steep spectrum measured by ROSAT could possibly represent intrinsic
spectral variability possibly due to the high energy end of the synchrotron
component, extending up to the soft X-ray band.

\subsection{2251+158 (3C454.4)}

As shown in Table 3 a fit with a power-law model and Galactic absorption
gives an unacceptable fit to the data. Using a free value for $N_H$ the
fit improves significantly, requiring a column density higher than the
Galactic value at the 99\% confidence level (see the contour plot in
Fig.4). The value of the required absorption column density in the quasar
rest-frame is $N_{H, int}=4.9^{7.9}_{2.4} \times 10^{21}$ cm$^{-2}$. 

An acceptable fit ($\chi_r=0.93$) of the low-energy portion of the
spectrum can also be obtained with a broken power-law model (and Galactic
column density), with a very flat ($\Gamma =-0.25$) soft spectrum. This,
as in the case of other sources (e.g. 0836-710, Paper I), could be due to
an intrinsic break in the continuum. From a statistical point of view
both models have almost the same probability.

\section{Modelling the SEDs}

Using $Beppo$SAX and historical data we assembled the SEDs of the
sources, shown in Fig.5. EGRET data are the average over the available
observations as given by Hartman et al. (1999), while upper limits are
from Thompson et al. (1995). For 0208-512 we report also the COMPTEL
detection reported in Blom et al. (1995).

In the IR-optical bands, for 2223-052 are available quasi-simultaneous
ISO (reported in Haas et al. 1998) and optical data.For other 3
sources (1641+399, 2243-123, 2251+158) we report average optical fluxes
with the observed variability range. For 1641+399 Xie et al. (1999)
report a variability amplitude in the $B$-band $\Delta B\simeq 3$ mag
(which is shown in Fig. 5), while for the other 2 sources the variability
observed is rather small (less than 1 mag.). For 0208-512 and 0521-365 we
report only the data we could find in literature. Both sources have IUE data.
References to the data are reported in the figure caption.

The double-humped shape of the Spectral Energy Distribution (SEDs) is
widely interpreted as due to synchrotron-Inverse Compton emission.  The
low energy component is synchrotron radiation from a population of
relativistic electrons, while the high energy component is believed to be
produced through Inverse Compton scattering of soft photons by the same
electrons (although other scenarios have been considered, e.g. hadronic
models, see Mannheim \& Biermann 1992). The nature of soft photons
involved in the IC process is still debated. In Blazars where thermal
features (emission lines, the {\it Blue Bump}) are weak or even absent
(as in BL Lac objects) the energy density of target photons is probably
dominated by those produced by Synchrotron (SSC model, e.g. Maraschi,
Ghisellini \& Celotti 1992), while in powerful quasar (FSRQ) photons
coming into the jet from the external environment form the dominant
population (EC model, Dermer \& Schlickeiser 1993; Sikora, Begelman \&
Rees 1994).

We have reproduced the overall SEDs using a Synchrotron-Inverse Compton
(SSC+EC) model. The source is modeled as a sphere with radius $R$,
tangled magnetic field with intensity $B$, in motion with bulk Lorentz
factor $\Gamma _{\rm b} $ at an angle $\theta $ with respect to the line
of sight. Relativistic effects are then regulated by the relativistic
Doppler factor, given by: $\delta =[\Gamma _{\rm b}(1-\beta \cos
\theta)]^{-1}$. Usually for Blazars $\theta \sim 1/\Gamma _{\rm b} $,
implying $\delta\simeq \Gamma _{\rm b} $. Relativistic electrons emit
through synchrotron and IC mechanisms. IC emissivity is calculated taking
into account the full Klein-Nishina cross-section (following the Jones
1968 treatment).  The electron energy distribution is modeled with the
form:
\begin{equation} 
N(\gamma )=K\gamma ^{-n_1}\left( 1+ \frac{\gamma }{\gamma _b
}\right)^{n_1-n_2}
\end{equation} 
\noindent 
where $K$ is a normalization factor, $\gamma _b$ is the break Lorentz
factor, $n_1$ and $n_2$ are the spectral indices below and above the
break, respectively. This particular form for the distribution function
has been assumed on a purely phenomenological basis, in order to describe
the curved shape of the SED.

The external radiation field is modelled as a black-body peaking at $\nu
_{\rm ext}\simeq 10^{15}$ Hz and energy density $U_{\rm ext}$. The last
quantity is calculated assuming that the radiation, with total luminosity
$\tau L_{\rm ext}$ ($\tau $, usually assumed to be $\sim 0.1$ is the
fraction of the central emission reprocessed by the Broad Line Region) is
diluted into a sphere of radius $R_{\rm BLR}$. For 2251+158 which shows a
clear UV bump we take $L_{\rm ext}=L_{\rm UV}$, while for the other cases
we inferred $L_{\rm ext}$ from the luminosity of broad emission lines,
following the approach used by Celotti et al. (1997). Luminosities are
reported in Table 5. The value of $R_{\rm BLR}$ has been choosen to fit
the observed data, but we checked that it is close to the value expected
from the relation between the luminosity of the BLR and the radius
derived (for radio-quiet quasars) by Kaspi et al. (2000). The external
radiation energy density is amplified in the blob's rest frame by a
factor $\Gamma ^2$ (e.g. Ghisellini et al. 1998). Moreover, as pointed
out by Dermer (1995) (and confirmed by Georganopoulos, Kirk \&
Mastichiadis 2001), because of
relativistic effects the beaming pattern of the EC radiation is narrower
than that of the SSC emission. However, when $\delta\simeq \Gamma$ (as
assumed here), the amplification is the same for both processes.

With these assumptions the model is well constrained by the observation
and we can obtain fairly robust estimates of the principal physical
quantities in the jet (see Paper I). The values of the parameters used to
reproduce the SEDs are reported in Table 5. In all cases the bulk of the
IC component is reproduced with the EC emission, while in some cases the
SSC spectrum (peaking at lower energies) partly accounts for the low
energy portion of the X-ray spectrum. The cut-off in the synchrotron
spectrum around $10^{11}$ Hz is due to self-absorption. Clearly the
adopted one-zone model can not account for the radio emission, widely
considered to be due to the superposition of the emission produced in the
outer ($d \gta 0.1$ pc) regions of the jet (e.g. Blandford \& Konigl
1979).

\section{Discussion and Conclusions}

We have presented the results of the analysis of 6 FSRQs observed with
$Beppo$SAX. The smooth power-law ($\Gamma =1.6-1.7$) X-ray continuum is
produced through Inverse Compton scattering on photon external to the
jet, except for the ``atypical'' source 0521-365. In the case of two
sources, 0208-512 and 2251+158, there is evidence of more complex X-ray
spectra. In particular 0208-512 shows an absorption feature at $\sim 0.6$
keV, that we suggest could be explained allowing the presence of a warm
absorption intercepting the emission produced in the jet.

The soft X-ray spectrum of 2251+158 shows a clear deficit of photons,
consistent either with an intrinsic absorption (with rest-frame column
density $\sim 5\times 10^{21}$ cm$^{-2}$) or an intrinsic break in the
continuum occurring below $\sim $1 keV. In the latter case the position of
the break could provide important insights into the physics of the jet,
in particular regarding the value of the minimum energy of the radiating
particles. On the other hand there is growing evidence that the presence
of intrinsic absorption is common in high-redshift radio-loud quasars
(e.g., Reeves \& Turner 2000). Indeed $Beppo$SAX observations of two
distant ($z\sim 4$) blazars (Fabian et al. 2001a,b) show the presence of
huge absorption, with column density of the order of few $10^{23}$
cm$^{-2}$. 

For three sources (0208-512, 1641+399 and 2251+158) we detect hard X-ray
emission up to $\sim 100$ keV. For 1641+399 and 2251+158 the hard X-ray
component appears to be the smooth extrapolation of the soft-medium X-ray
continuum, while in 0208-512 the PDS spectrum is likely contaminated by
another source present in the large FOV of the instrument. We note
however that 0208-512 is one of the few blazars detected by COMPTEL (Blom
et al. 1995) and because of its exceptionally intense flux in the MeV
band is considered as a prototype of the so-called ``MeV
Blazars''. Therefore we can not completely exclude the possibility that
the spectral component responsible for the MeV flux contributes also in
the PDS band, explaining the observed excess.

The modelling of the SEDs shows that FSRQs (except the case 0521-365, see
below) are characterized by jets with Doppler beaming factors larger than
10 and magnetic field of the order of $B\sim 2-3$ G, roughly in
equipartition with the emitting electrons. The size of the emitting
region is of the order of $10^{16}$ cm, consistent with the absence of
short-timescale variability ($\lta 1$ day) in this class of blazars. These
results confirm, with a larger sample, the conclusions of Paper I

An interesting issue for Blazars is the value of the minimum Lorentz
factor of the emitting electrons, $\gamma _{\rm min}$. In Paper I we
showed how, from the absence of a spectral break in the soft X-ray
continuum, it is possible to infer that the electron energy distribution
extends down to $\gamma _{\rm min} \sim 1$. In fact a break in the EC
continuum is expected at the frequency $\nu _{\rm br}\simeq \Gamma ^2 \nu
_{\rm ext} \gamma _{\rm min} \sim 10^{17} \Gamma _{1}\sim $ (e.g. Sikora
et al. 1997), corresponding to an energy $\sim 1$ keV (in the rest frame
of the source). We recall that the lack of soft X-ray photons observed in
2251+158 (as for the case of 0836+710 reported in Paper I) could be
interpreted as due to such intrinsic break in the continuum. Observations
with larger signal to noise (possible with $XMM-Newton$ or $Chandra$)
will allow to disentangle the two possible interpretations, intrinsic
curved continuum or absorption.

In our fits $n_1$, the index of the low energy portion of the electron
energy distribution, is smaller than 2. Clearly such a flat spectrum can
not be produced by the cooling of high energy electrons (in that case ine
would expect $n_1=2$). On the other hand this spectrum is even flatter
than the standard prediction $n_1=2$ for a population of non-thermal
electrons produced through Fermi acceleration by a non-relativistic shock
(e.g. Blandford \& Ostriker 1978). However recent calculations (see the
review by Kirk \& Dendy 2001) show that relativistic shocks, such as those
likely present in the jet of blazars, could produce extremely flat
($n_1< 2$) spectra.

The case of 0521-365 is atypical. Previous observational evidence
suggested that the jet forms a relatively large angle with the line of
sight. Pian et al. (1996), using independent arguments (in particular the
presence of an optical jet,e.g. Scarpa et al. 1999), concluded that
$\theta \sim 30^{\rm o }$. Assuming that the bulk Lorentz factor of the
emitting plasma is similar to that of the other blazars, $\Gamma _{\rm b}
\sim 10$, this implies that the emission from 0521-365 is weakly boosted
($\delta =1-2$). The SED reported in Fig.5 has been calculated assuming
$\Gamma _{\rm b} =10$ and $\theta =15^{\rm o }$, implying $\delta=3$. In
this situation the X-ray/$\gamma$-ray continuum, contrary to the case of
the other sources, is dominated by the SSC emission, while the EC
spectrum gives only a small contribution in the $\gamma $-ray band.

Part of the motivation for this investigation was to study the difference
between the EGRET and non-EGRET sources. In the whole FSRQ sample
observed with \sax there are 9 EGRET sources and 3 sources not detected
by EGRET. The average X-ray spectral indices ($\Gamma _{\rm EGRET}=1.56
\pm 0.06$ and $\Gamma _{\rm non-EGRET}= 1.69\pm 0.07$ for EGRET and
no-EGRET sources, respectively) indicate that, within the statistical
uncertainty, the X-ray spectral characteristics are similar for both
groups. The modelling of the SEDS confirms the strong similarity between
the two groups.  The lack of the EGRET detection could be attributed to
the variability behaviour typical of these sources in the $\gamma $-ray
band (e.g. Mukherjee et al. 1997) rather than to an intrinsic difference
in the properties of the jet. New $\gamma $-ray missions (AGILE, GLAST)
will help to further investigate these problems.

\acknowledgments{We thank the $Beppo$SAX SDC for providing us with the
cleaned data. This work was partly supported by the ASI grant
I-R-105-27-00. This research has made use of the NASA/IPAC Extragalactic
Database (NED) which is operated by the Jet Propulsion Laboratory,
California Institute of Technology, under contract with the National
Aeronautics and Space Administration.}

\clearpage

\vskip 1.5 truecm

\centerline{ \bf Figure Captions}

\vskip 1 truecm

\figcaption[]{Fit of LECS and MECS data of 0208-512 with a power-law and
free absorption. The feature at $\sim 0.6$ keV is clearly visible in the
data to model ratio (lower panel).\label{0208_noline}}

\figcaption[]{Fit of LECS and MECS data of 0208-512 with a power-law
(Galactic absorption) plus an absorption
edge.\label{0208_line_edge}}


\figcaption[]{Fit of LECS data of 1641+399 with a power-law and
Galactic absorption. \label{1641_pl}}

\figcaption[]{68, 90 and 99\% photon index-$N_H$ confidence contours for
2251+158. The column density exceeds the Galactic value at $>$99\%
level.\label{0208_cont}}

\figcaption[]{Spectral Energy Distributions of the sources analyzed in
this work. The solid line is the model discussed in the text. References
for the data: {\bf 0208-512}: Bertsch et al. 1993, Tornikoski et
al. 1996, Stacy et al. 96, Impey \& Tapia 1988, Blom et al. 1995 {\bf
0521-365}: Wright \& Otrupcek 1990, Wright et al. 1996, Glass et al. 1981
, Falomo et al. 1993, Xie et al. 1998. {\bf 1641+339}: Wiren et al. 1992,
Patnaik et al. 1992, Geogory \& Condon 1991, Moshir et al. 1990,
Neugebauer et al. 1982, Villata et al. 1997, Xie et al. 1998. {\bf
2223-052}: Wright \& Otreepcek 1990, Haas et al.  1998, OPTICAL. {\bf
2243-123}: Wright \& Otreepcek 1990, Wright et al. 1983, Kotilainen et
al. 1998, Garcia et al. 1999. {\bf 2251+158}: Kuhr et al. 1981, NED, Gear
et al. 1994, Stevens et al 1994, Impey \& Neugebauer 1988, Smith et
al. 1988, Pian \& Treves 1993. \label{SED}}

\begin{table}
\begin{center}
\caption{List of the sources discussed in the present work. The last column
indicates EGRET detection or lack thereof.}
\begin{tabular}{lccl}
\hline
Source & Other name& $z$ & EGRET\\
&&&\\
\hline
0208-512 &PKS & 1.003& Y\\
0521-365 & PKS& 0.055&  Y\\
1641+399 &3C 345 &0.593& N\\
2223-052 &3C 446 &1.4&  N\\
2243-123 & PKS &0.63&  N\\
2251+158 &3C 454.4 & 0.859 & Y \\
\hline
\end{tabular}
\end{center}
\end{table}
\newpage

{\small
\begin{table*}
\begin{center}
\caption{{\it Beppo}SAX data observation log.}
\hspace*{-1.5cm}
\begin{tabular}{lcccccccc}
\\
\hline
\hline
Date & Start& End& LECS& net cts/s$^a$& MECS &net cts/s$^b$ & PDS& net cts/s\\
 & & &Exp.(s) & &Exp.(s) & & Exp.(s)&\\
\hline
&&&&&&&&\\
\multicolumn{9}{c}{\bf 0208-512} \\ 
\hline
14-15/01/01 &13:37:10 & 13:00:09 & 15650 & $0.022 \pm 0.001$ & 34256 &$0.053\pm 0.001$& 15260& $0.1417\pm0.03$ \\
\multicolumn{9}{c}{\bf 0521-365} \\ 
\hline
2-3/10/98 & 10:16:24 & 09:46:23& 17752& $0.067 \pm 0.002$ &41082 &$0.106 \pm 0.001$ & 19052& $0.05 \pm 0.04$\\
\multicolumn{9}{c}{\bf 1641+399} \\ 
\hline
19/02/99 & 03:55:43 & 20:13:57 & 10934 & $0.045 \pm 0.002$ & 25866 &$0.059 \pm 0.002$ & 13052& $0.15\pm 0.05$\\
\multicolumn{9}{c}{\bf 2223-052} \\ 
\hline
10-11/11/97 & 13:22:27 & 00:22:40 & 9200 & $ 0.01\pm 0.001$ & 16180 &$0.016 \pm 0.001$ & 6860 & $ 0.008\pm $ 0.08\\
\multicolumn{9}{c}{\bf 2243-123} \\ 
\hline
18-19/11/98 & 18:41:31 & '11:49:58 & 10060 & $0.015\pm0.002$ & 27500 & $0.022\pm0.001$ & 13222 & $0.004 \pm 0.015$\\
\multicolumn{9}{c}{\bf 2251+158} \\ 
\hline
5-6/6/2000 & 15:36:05 & 21:44:38 & 17425 & $0.046\pm0.001$ & 48515 &$ 0.115\pm 0.002$ & 22439 & $0.29 \pm 0.03$\\
\hline
\multicolumn{9}{l}{$^a$: 0.1-4 keV} \\
\multicolumn{9}{l}{$^b$: 1.8-10.5 keV, 2 MECS units} \\
\end{tabular}
\end{center}
\end{table*}
}

\begin{table*}
\begin{center}
\caption{Absorbed power-law fits to {\it Beppo}SAX data. Errors are quoted
at the 90\% confidence level for 1 parameter of interest.}
\begin{tabular}{lcccc}
\\
\hline
\hline
Source & \, \, $\Gamma ^a $\, \, & $N_H$ & $F_{[2-10\, \rm keV]}$ & $\chi^2/$d.o.f.\\ 
      & & 10$^{20}$ cm$^{-2}$ & $10^{-12}$ erg cm$^{-2}$ s$^{-1}$  &  \\
\hline
\hline
{\bf 0208-512} (LE+ME2+PDS)$^b$ & $1.68 ^{1.80}_{1.51}$ & $16.7^{30.0}_{0.1} $ & 4.7 & 51.16/28 \\
\hline
{\bf 0521-365} (LE+ME) &$1.77 ^{1.85}_{1.73}$ & $4.3_{3.2}^{7.6} $& 8.72 & 69.35/89\\
\hline
{\bf 1641+399}  (LE+ME+PDS) &$1.57_{1.47}^{1.66}$& 1.12(fix) & 5.2  & 65.26/45 \\
\hline
{\bf 2223-052} (LE+ME) &$1.84 ^{2.27}_{1.56}$ &$7.1_{0}^{76.1} $ &1.25  & 14./16\\
-&$1.86 ^{2.15}_{1.58}$ &4.87 (fix) & -& 14.72/17\\
\hline
{\bf 2243-123} (LE+ME) &$1.73 ^{1.91}_{1.56}$& $3.05_{1.28}^{7.7}$ & 1.83 &18.37/23 \\
\hline
{\bf 2251+158} (LE+ME+PDS) &$1.31 ^{1.26}_{1.36}$ &6.5(fix) &10.6  &94.2/80 \\
-&$1.38 ^{1.33}_{1.44}$ & $18.6 ^{25.1}_{12.2}$ &- & 77.1/79\\
\hline
\multicolumn{4}{l}{$^a$: photon index, related to the spectral index by $\alpha =\Gamma+1$.} \\
\multicolumn{4}{l}{$^b$: see Table 4}
\end{tabular}
\end{center}
\end{table*}

\begin{table*}
\begin{center}
\caption{Fits to {\it Beppo}SAX data of 0208-512. Errors are quoted at
the 90\% confidence level for 1 parameter of interest.} 
\begin{tabular}{lcccc}
\\
\hline
\hline
Model & $\Gamma^a$ & $E_{\rm abs}/N_H ^b$ & $\tau/\xi^c$ &  $\chi^2/$d.o.f.\\ 
\hline
\hline
PL+edge & $1.72^{1.88}_{1.65}$& $0.56^{0.75}_{0.33}$& $2.1_{0.9}^{10}$& 13.25/17\\
\hline
PL+notch & $1.68^{1.81}_{1.55}$& $0.89^{0.94}_{0.80}$&
$0.21^{0.31}_{0.13}$&12.85/17  \\
\hline
PL+ABSORI & $1.70^{1.96}_{1.62}$& $4.2^{10.3}_{1.2}$& $262^{853}_{32}$&
15.13/17\\ 
\hline
\multicolumn{5}{l}{$^a$: photon index, related to the spectral index by
$\alpha =\Gamma+1$.} \\ 
\multicolumn{5}{l}{$^b$: Energy of the absorption feature (keV) or $N_H$
for ABSORI (in units of $10^{20}$ cm$^{-2}$).} \\
\multicolumn{5}{l}{$^c$: optical depth for the edge, width (keV) for
NOTCH or $\xi$ for ABSORI} \\
\end{tabular}
\end{center}
\end{table*}

\begin{table*}
\begin{center}
\caption{Parameters used for the emission model described in the text.}
\begin{tabular}{ccccccccc}
\\ 
\hline 
\hline 
$R$ & B & $\delta$ & $\gamma _{\rm b}$ & n$_1$& $n_2$ & $K$ & $L_{\rm BLR}$ & $R_{\rm BLR}$\\
$10^{16}$ cm& G& & & & & cm$^{-3}$ & $10^{45}$ erg s$^{-1}$ &$10^{18}$ cm \\ 
\hline
&&&&&&&&\\
\multicolumn{9}{c}{\bf 0208-512} \\
\hline
1.5& 1.5 & 18& 100& 1.4& 3.8& $2\times 10^4$& 2.1& 0.6\\

\multicolumn{9}{c}{\bf 0521-365} \\
\hline
2.0& 0.3& 3$^*$ & $8.8\times 10^3$  &  1.25& 4 &$3\times 10^3$ & $2.4\times
10^{-2}$ & 0.8 \\

\multicolumn{9}{c}{\bf 1641+399} \\
\hline
4.0& 2.9& 9.75& 200& 1.5& 4.2& $2.8\times 10^3$& 3.7& 0.6\\

\multicolumn{9}{c}{\bf 2223-052} \\
\hline
4.25 & 5.6& 17& 135& 1.6& 4.3& $1.7\times 10^3$& 8.5& 1.1\\

\multicolumn{9}{c}{\bf 2243-123} \\
\hline
3.5 & 2.5& 15& 250& 1.6& 4.3& $1.7\times 10^3$& 5.6&0.9 \\

\multicolumn{9}{c}{\bf 2251+158} \\
\hline
4.0 & 1.5& 12& 60& 1.8& 3.4& $5\times 10^4$ & 4. & 1.0\\
\hline
\multicolumn{9}{l}{$^*$: see text} \\
\end{tabular}
\end{center}
\end{table*}

\newpage

\begin{figure}
\centerline{\plotone{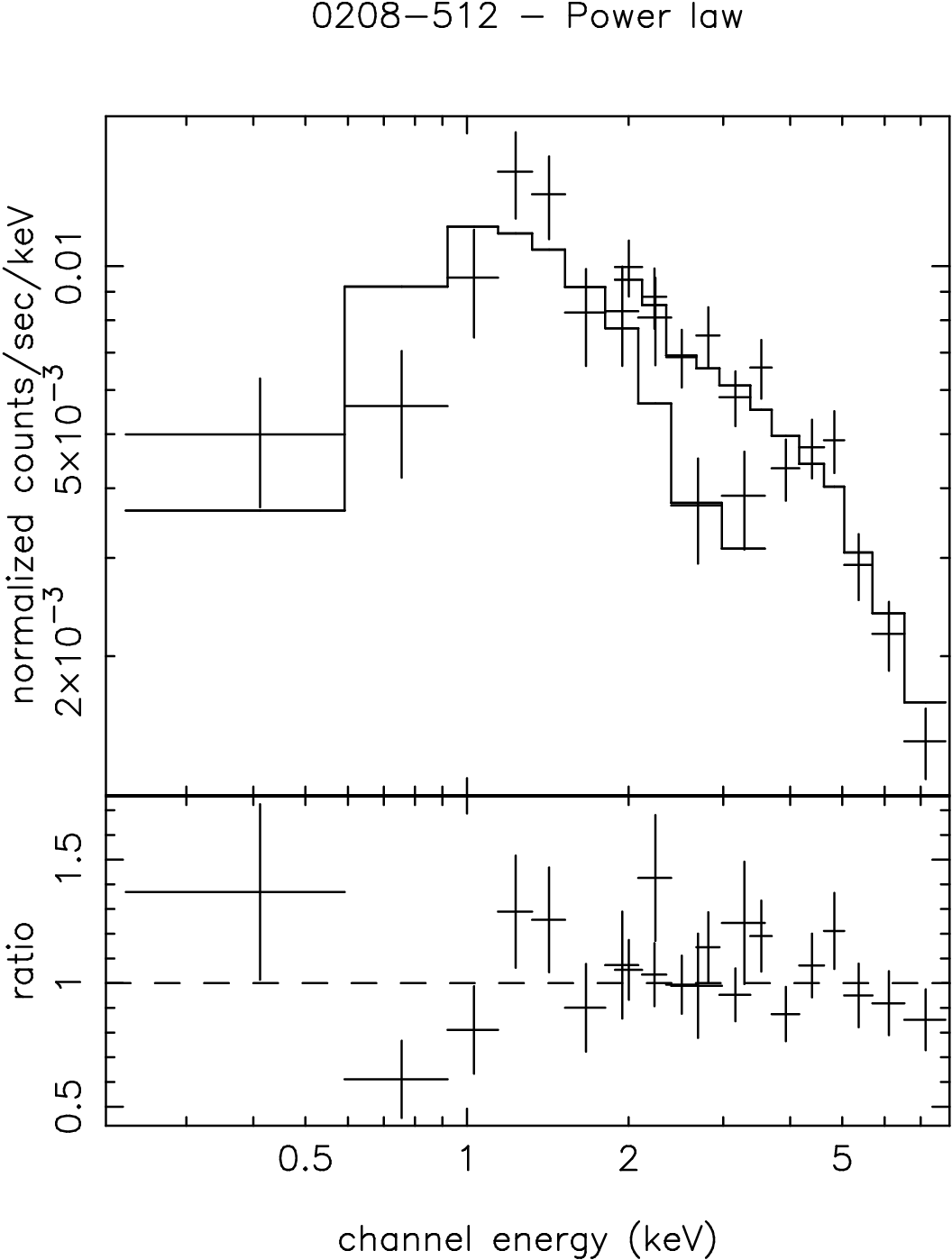}}
\end{figure}
\clearpage

\begin{figure}
\centerline{\plotone{tavecchio_fig2.ps}}
\end{figure}
\clearpage

\begin{figure}
\centerline{\plotone{tavecchio_fig3.ps}}
\end{figure}
\clearpage

\begin{figure}
\centerline{\plotone{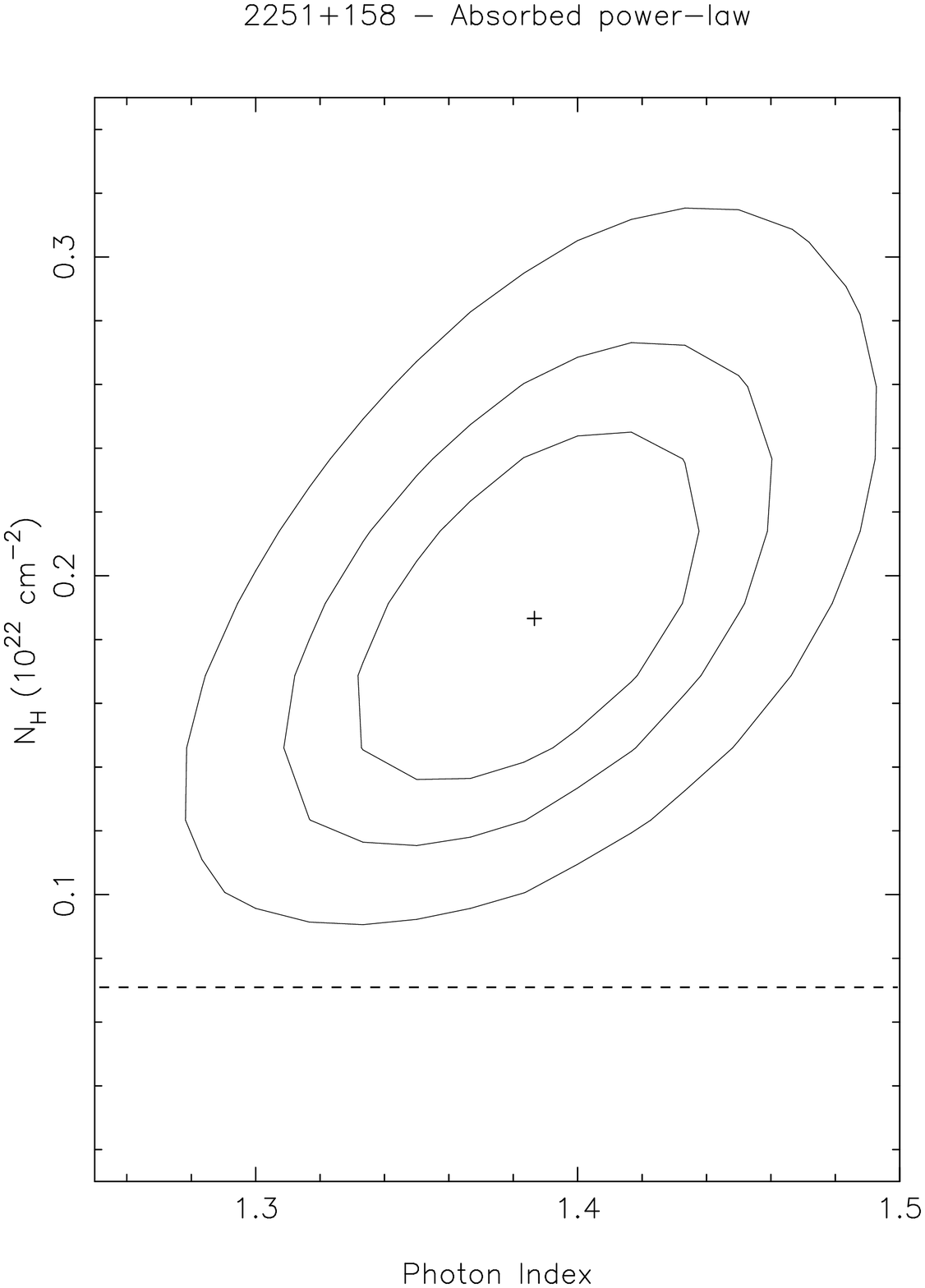}}
\end{figure}
\clearpage

\begin{figure}
\centerline{\plotone{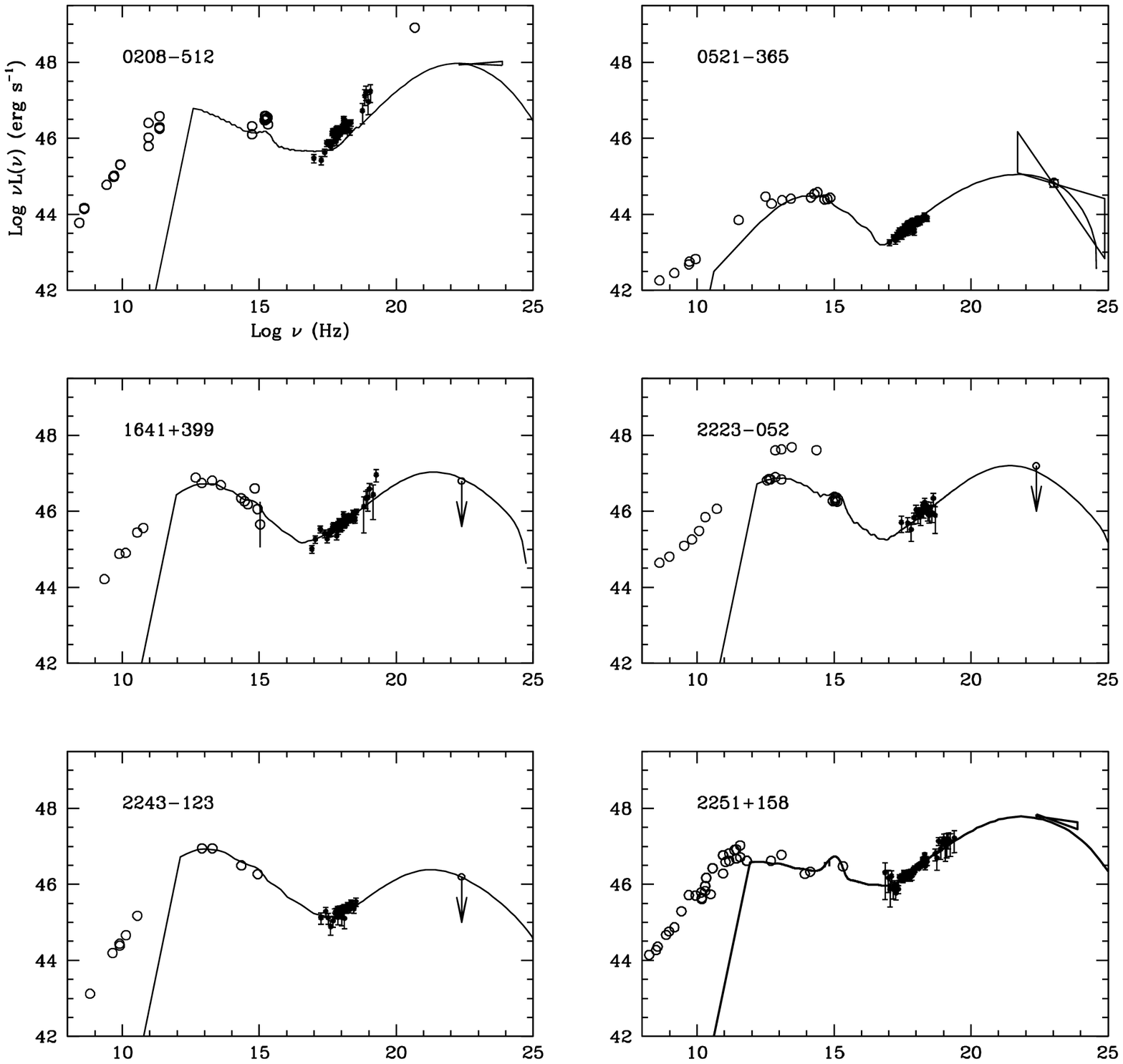}}
\end{figure}
\clearpage

\end{document}